\documentclass[]{spie}  %>>> use for US letter paper
\usepackage[]{graphicx}
\usepackage{epstopdf}
\usepackage{amssymb}

\title{Fourier domain asymmetric cryptosystem for privacy protected 
multimodal biometric security}

\author{Debesh Choudhury   %\supit{a} and Barry B. Author2\supit{b}
\skiplinehalf
 Neotia Institute of Technology, Management and Science\\
 Department of Electronics and Communication Engineering\\ D. H. Road, 
 Jhinga, PO - Amira, South 24 Parganas\\ Pin - 743368, West Bengal, India
}

\authorinfo{Further author information: E-mail: 
debesh@iitbombay.org, Tel: +91-9831369809}

\begin{document} 
  \maketitle 

%%%%%%%%%%%%%%%%%%%%%%%%%%%%%%%%%%%%%%%%%%%%%%%%%%%%%%%%%%%%% 
 \begin{abstract}
 We propose a Fourier domain asymmetric cryptosystem for multimodal 
biometric security. One modality of biometrics (such as face) is used as 
the plaintext, which is encrypted by another modality of biometrics 
(such as fingerprint). A private key is synthesized from the encrypted 
biometric signature by complex spatial Fourier processing. The encrypted 
biometric signature is further encrypted by other biometric modalities, 
and the corresponding private keys are synthesized. The resulting 
biometric signature is privacy protected since the encryption keys are 
provided by the human, and hence those are private keys. Moreover, the 
decryption keys are synthesized using those private encryption keys. The 
encrypted signatures are decrypted using the synthesized private keys 
and inverse complex spatial Fourier processing. Computer simulations 
demonstrate the feasibility of the technique proposed.
 \end{abstract}


%%%%%%%%%%%%%%%%%%%%%%%%%%%%%%%%%%%%%%%%%%%%%%%%%%%%%%%%%%%%%

\section{BACKGROUND of Cryptography}
 \label{sec:background}

Secure communication and sharing of secret information often require 
special expensive channels. Alternatively, the information may be 
encrypted and may be exchanged between the users through public 
channels. In old days, the secret keys used to be distributed through 
secure channels, generally by registered post or courier, which are slow 
and costly. This is known as symmetric key cryptography. The invention 
of public key cryptography by Diffie and Hellman~\cite{dh:land}, and 
independtly by Merkle~\cite{merkle} have radically changed the 
approaches of key distribution in cryptography. A public key 
cryptosystem~\cite{dh:land} has two keys, one for encryption and one for 
decryption. The two keys effect inverse operations and hence are 
related, but the decryption key can't be easily derived from the 
encryption key~\cite{dh:land}. The public key cryptography uses 
different keys for encryption and decryption, and hence is known as 
asymmetric key cryptography. The field has been continually enriched by 
the research works of Rivest {\it et al.}~\cite{RSA}, Merkle and 
Hellman~\cite{MH} and McEliece~\cite{McEliece}. The 
Rivest-Shamir-Adleman public key scheme or the widely known as RSA 
scheme rely on the fact that it is easy to generate two large primes and 
multiply them together, but it is very difficult to factor the 
result~\cite{RSA}. Merkel and Hellman have used the trapdoor knapsack 
problems which are also very difficult to solve~\cite{MH}. McEliece's 
cryptosystem is based on algebraic coding theory where a random Goppa 
code is chosen to synthesize the encryption key by permutation, the 
secret decryption key being the permutation and choice of Goppa code. In 
all these seminal cryptosystems, the secret decryption keys are required 
to be created from the public encryption keys following some secret 
mathematical rules or puzzles~\cite{hellman2002}.

\section{Motivation for Biometrics and Privacy Protection}
 \label{sec:Motiv and Intro}

Biometrics refer to the physiological human traits such as face, 
fingerprint, iris, palm prints etc. and the behavioral human traits such 
as gait, voice, walking style, gesture, typing rhythm etc. Biometric 
signatures are unique to individuals and hence are reliable for 
identifying human beings. Biometric authentication is gaining more 
acceptance day by day in a variety of applications in governmental 
programs as well as in personal verication 
problems~\cite{pb-book,bb-2009}. Biometrics can provide a way to 
synthesize keys for asymmetric encryption. Biometric data is also more 
difficult to be stolen or attacked making the encryption process 
inherently more secure and robust than storing text 
keys~\cite{uludag:2004}. Moreover, biometric keys being more secure, 
they also provide a way for personal identity verification because only 
the individual with matching biological traits would be able to produce 
the correct biometric key~\cite{jain:1999}.

A biometric template is a representation of several unique 
characteristics of individuals, and hence it contains sensitive personal 
and private information. Biometric identification requires a verifier to 
search for matches in the entire database of biometric signatures of the 
whole population. This can naturally open a possible security threat 
that the verifier can steal the biometric templates from the database. A 
malicious verifier can acquire the biometric templates and try severe 
impersonation attacks. Thus privacy protection is very essential in 
biometric authentication process because the biometric template data may 
be misused by dishonest verifiers~\cite{ratha:2001,spb:book2013}. 
Security and privacy are two essential requirements of biometric 
systems, but they mutually hinder each other. Biometric encryption has 
been an easy way for privacy protection and surmounting security 
threats~\cite{roberge:1999}, although there are several other methods 
available for securing the privacy information~\cite{spb:book2013}.

In this paper, it is tried to utilize the principles of asymmetric key 
cryptography to encrypt and decrypt biometric signatures using Fourier 
domain cryptography popularly known as double random phase 
encoding~\cite{javidi:PT} and phase truncated Fourier 
transform~\cite{peng:OL}. The encrypted biometric signatures are 
inherently privacy protected by the respective asymmetric keys derived 
from multimodal biometric traits of the humans.

\section{The proposed asymmetric cryptosystem}

The double random phase encoding cryptographic protocol relies on 
encryption of the plaintext using two random phase functions by two 
 \begin{figure}[b]
 \centering
 \includegraphics[width=10.5cm]{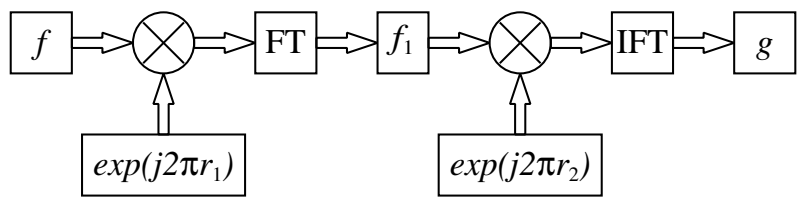} 
 \\[10pt]
  (a) \\[18pt]
 \includegraphics[width=10.5cm]{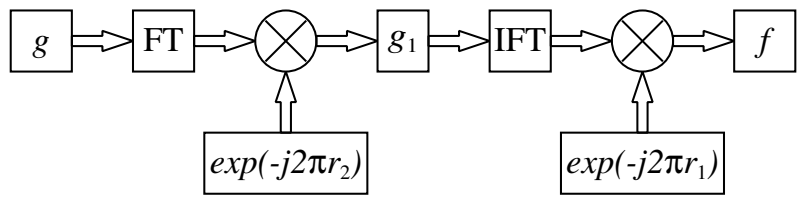} 
 \\[10pt]
 (b) \\[10pt]
 \caption{Double random phase encoding: (a)~encryption algorithm, and 
(b)~decryption algorithm.}  \label{drpe:scheme}
 \end{figure}
 Fourier transformations~\cite{javidi:PT,javidi:OL1995,goodman:FT}. The 
plaintext is multiplied by a random phase function (first private key) 
and Fourier transformed which is again multiplied by a second random 
phase function (second private key) and inverse Fourier transformed. 
Since this algorithm utilizes two random phase functions for encrypion, 
it is called double random phase encoding (DRPE). If we have an 
information specified by a spatial function $f(x,y)$, and have two 
independent white sequences $r_1(x,y)$ and $r_2(x,y)$, the encrypted 
function $g(x,y)$ after DRPE will be given by
  \begin{equation}
 g(x,y) = [ f(x,y) \exp \{j2\pi r_1(x,y)\}] \circledast \exp\{j2\pi 
r_2(x,y)\}
 \end{equation}

\noindent where $x,y$ are the space coordinates, $j=\sqrt{-1}$ and 
$\circledast$ is a convolution operation. The flow diagrams of 
encryption and decryption algorithms based on DRPE is given in 
Fig.\ref{drpe:scheme} where FT denotes a Fourier transform operator.

We propose an asymmetric cryptosystem that utilizes one modality of 
biometric signature as the plain text image, and another modality of the 
biometric signature is used as a private encryption key. The second 
biometric modality image is also encrypted using DRPE technique. The 
plain text image is multiplied by the complex exponential function of 
the DRPE encrypted second biometric signature, and the product is 
Fourier transformed. The phase part of the Fourier transform is 
extracted and is preserved as a private decryption 
key~\cite{peng:OL}. 
The amplitude part of the Fourier transform is utilized for further 
encryption by a third biometric signture. The product of the amplitude 
part of the Fourier
 \begin{figure}[htb]
 \centering
 \includegraphics[width=10.5cm]{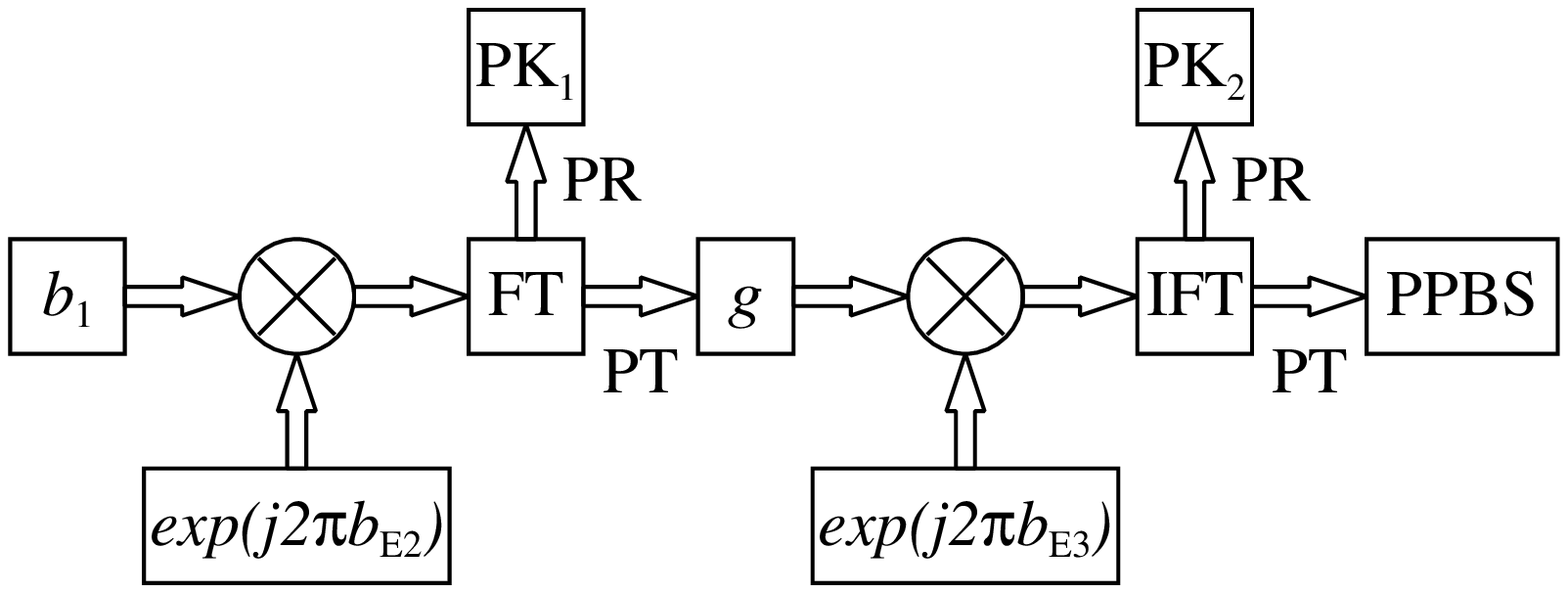} \\[10pt]
 (a) \\[18pt]
 \includegraphics[width=10.5cm]{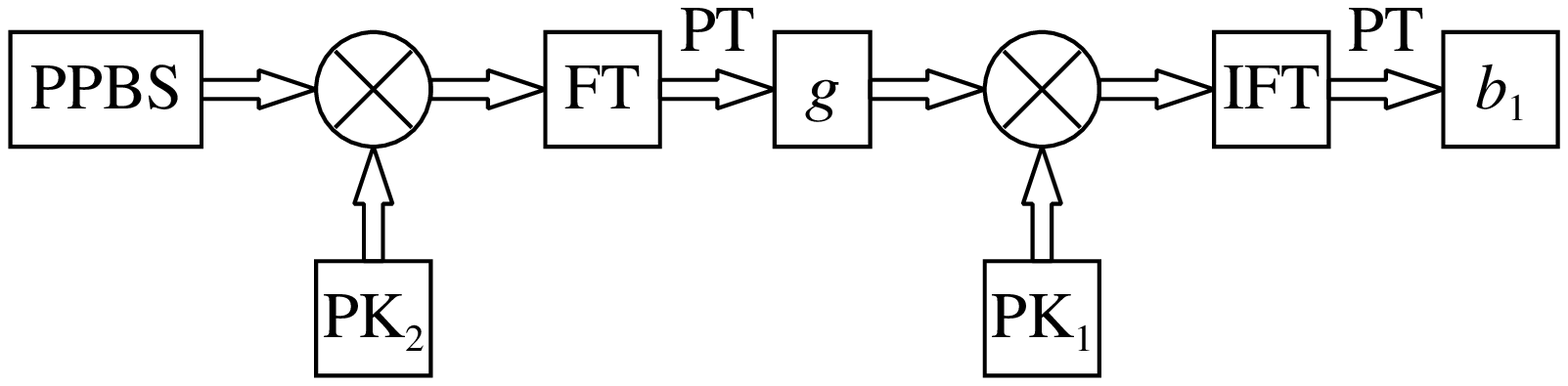} \\[10pt]
 (b) \\[10pt]
 \caption{The proposed asymmetric cryptosystem: (a)~encryption 
algorithm, and (b)~decryption algorithm.  \label{our_algo} }
 \end{figure}
 transform and the exponential function of the DRPE encrypted third 
biometric key is inverse Fourier transformed. Then, again the phase part 
of it is extracted and is preserved as a second private decryption key. 
The amplitude part of the Fourier transform is the encrypted privacy 
protected biometric signtaure.

Let $b_1$ be the first biometric trait image of a human (such as face), 
and $b_{E2}$, $b_{E3}$ be his/her other two biometric trait images (such 
as, fingerprint and iris) but are encrypted by DRPE. If PT be a phase 
truncation operator and PR be a phase reservation operation, we can 
express our cryptosystem using the flow diagram of 
Fig.\ref{our_algo}(a). The two private phase keys PK$_1$ and PK$_2$ are 
extracted by the PR operation. The final encrypted image PPBS represents 
the privacy protected biometric signature.

The encrypted biometric signature PPBS can be decrypted using the 
synthesized private phase keys PK$_1$ and PK$_2$ as shown in the 
flow diagram of Fig.\ref{our_algo}(b). These encrypted privacy protected 
biometric signatures can be preserved in the database. These data can be 
used for authentication by cross-correlation 
operation~\cite{vanderlugt}.

\section{Simulation Results}

\begin{figure}[htb]
 \centering
 \includegraphics[height=2.975cm]{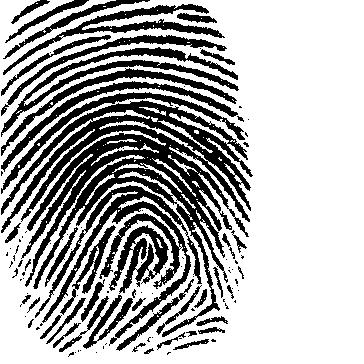}
 \includegraphics[height=2.975cm]{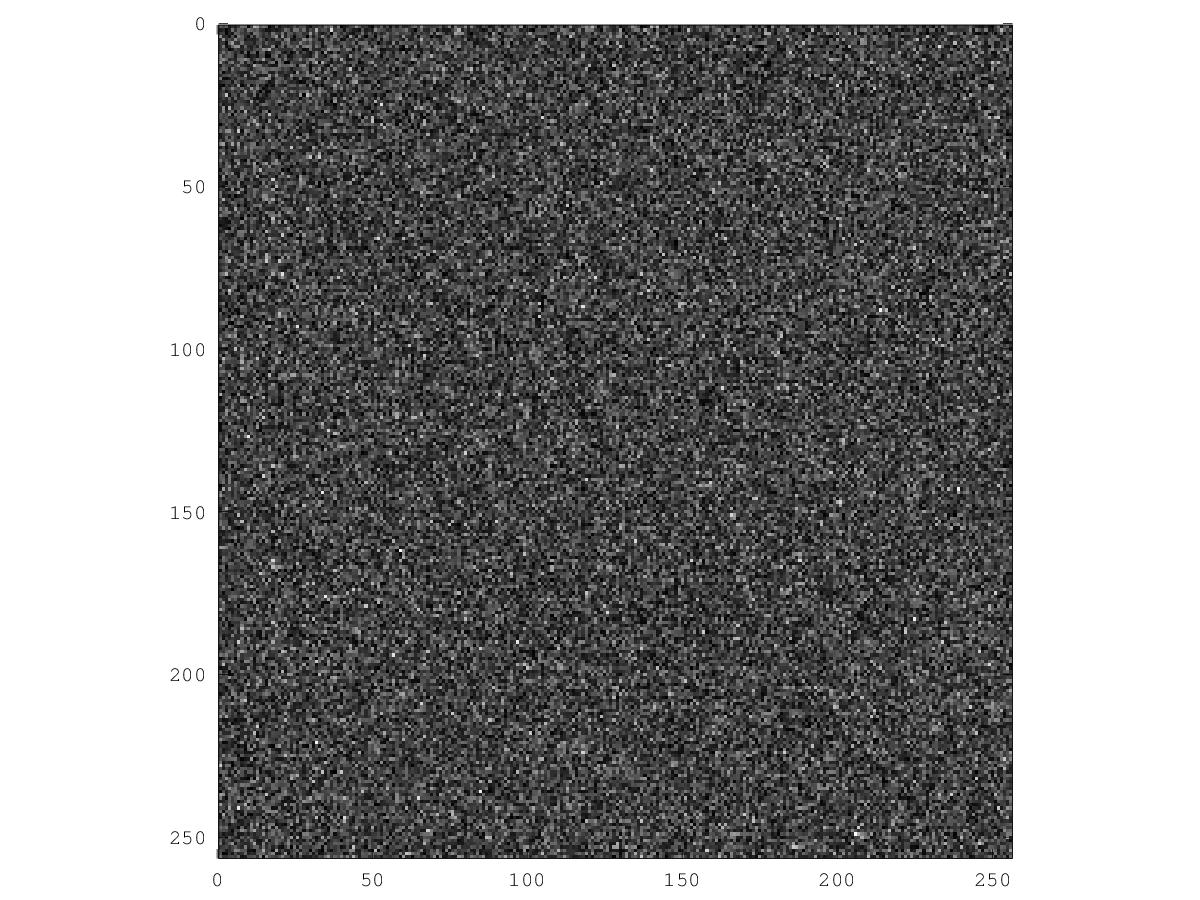}
 \includegraphics[height=2.975cm]{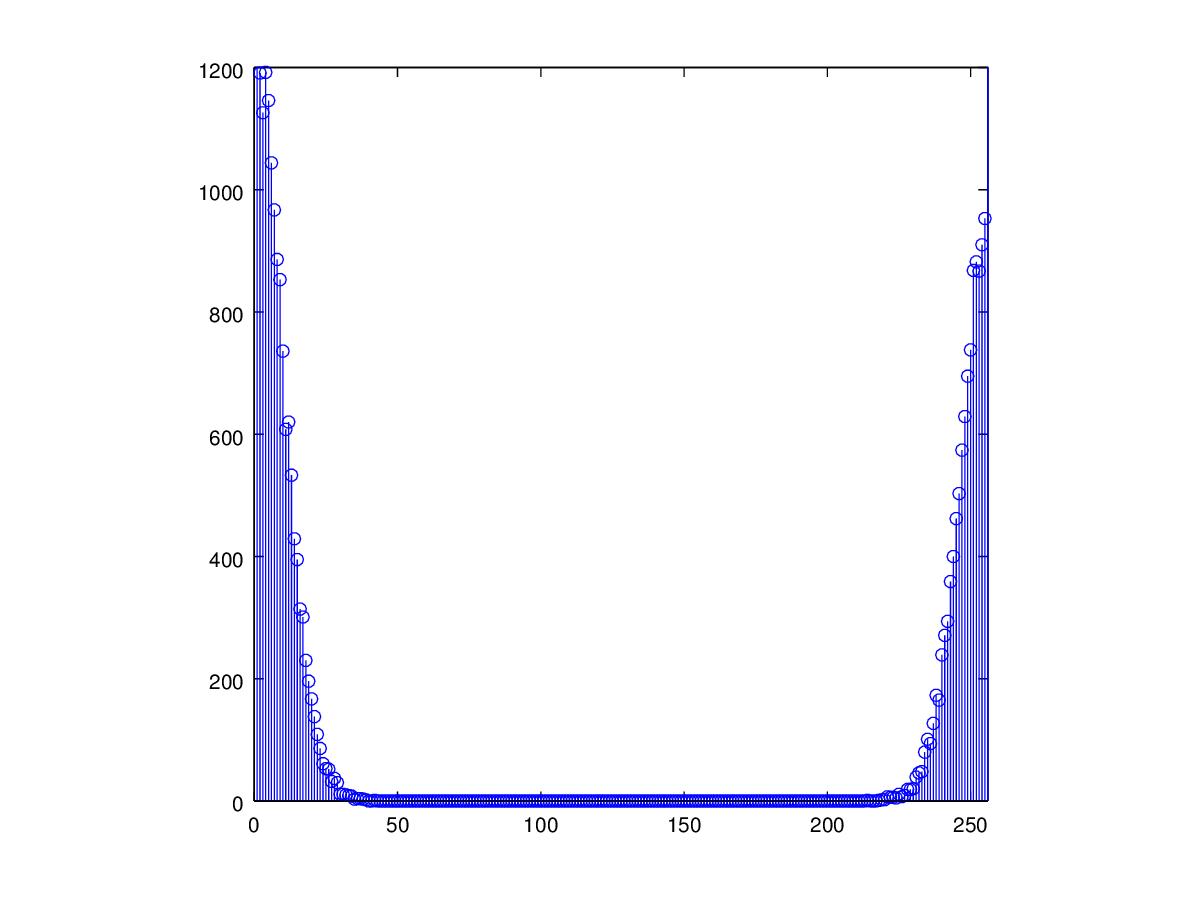}
 \includegraphics[height=2.975cm]{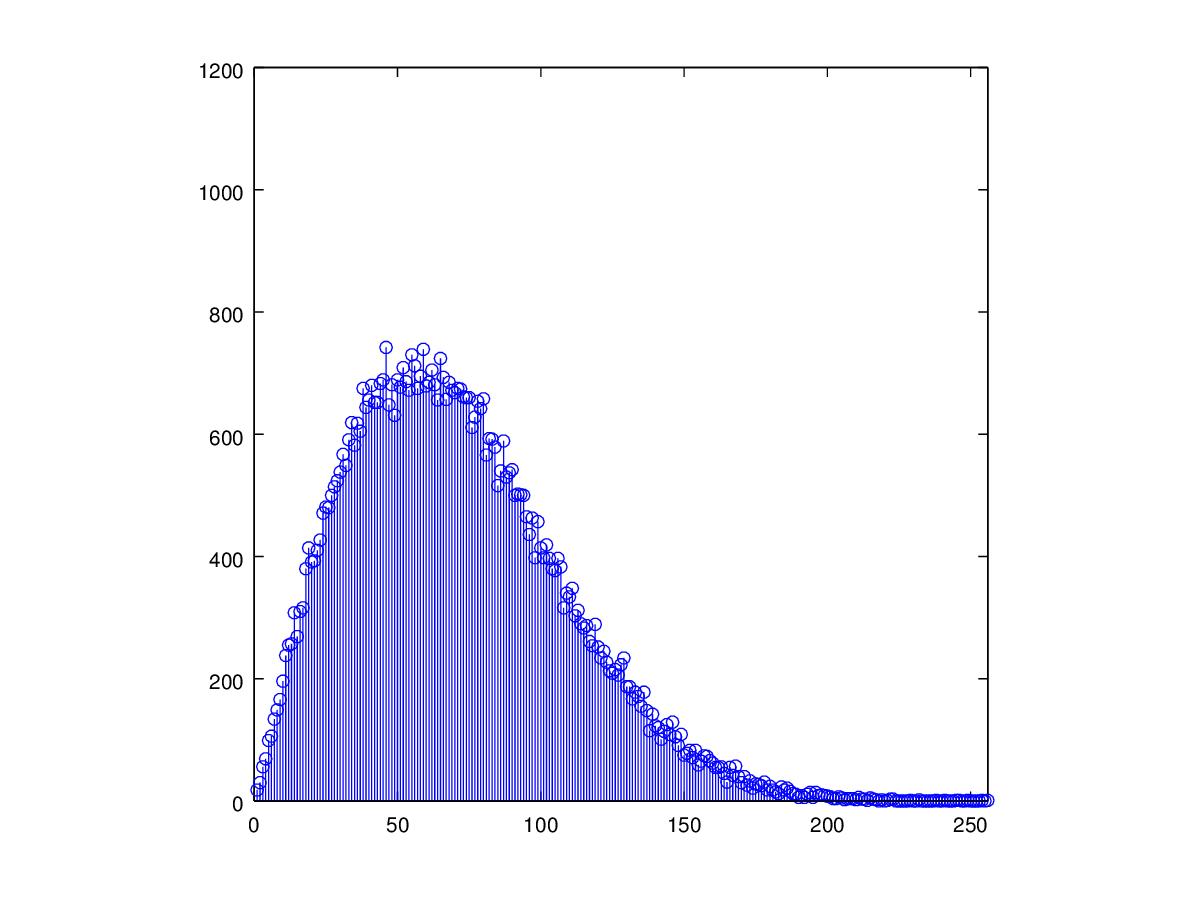} \\[5pt]
 \hglue -0.4in (a) \hskip 1.4in (b) \hskip 1.4in (c) \hskip 1.3in (d)\\[10pt]
 \includegraphics[height=2.975cm]{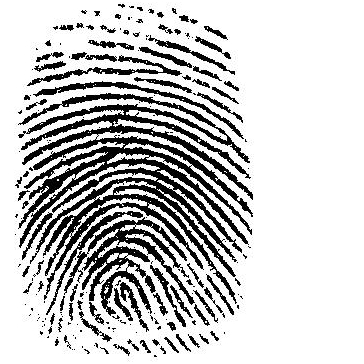}
 \includegraphics[height=2.975cm]{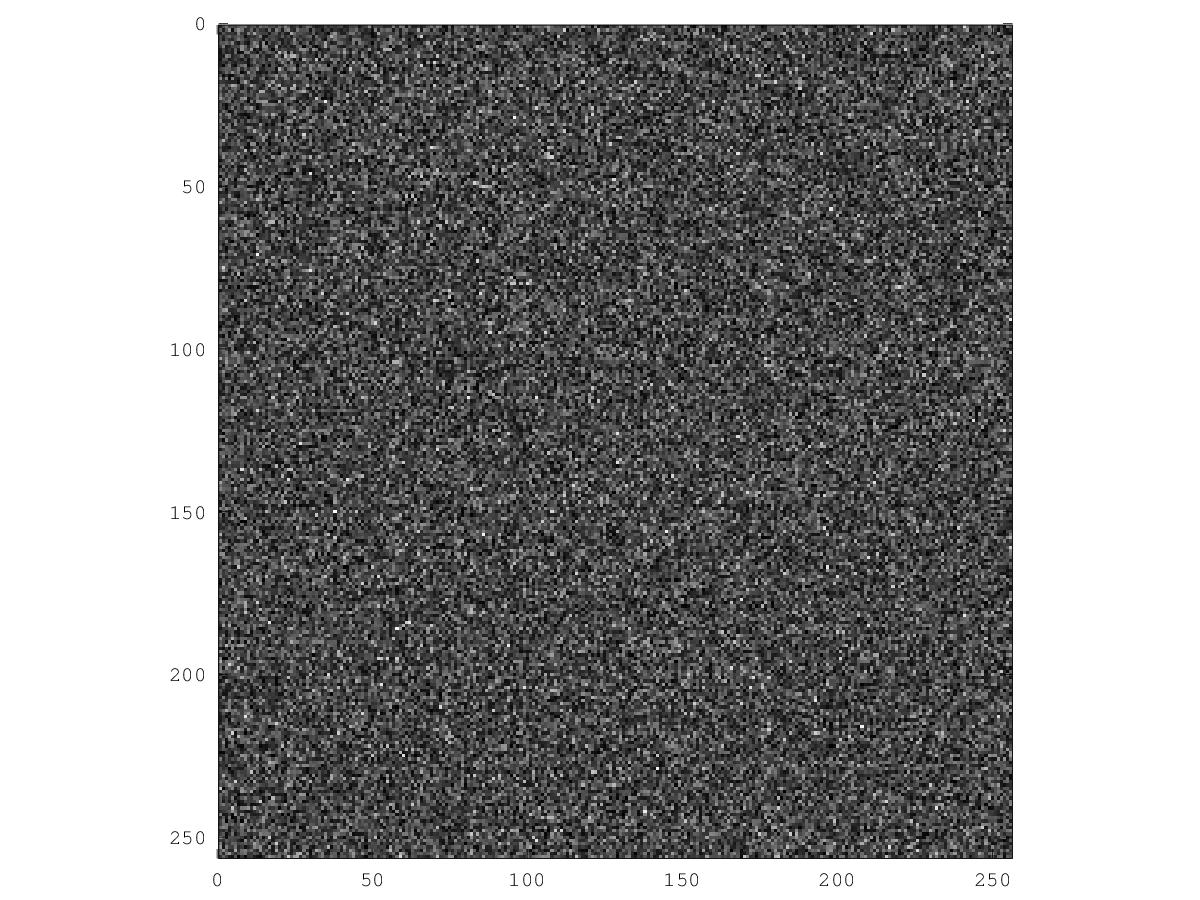}
 \includegraphics[height=2.975cm]{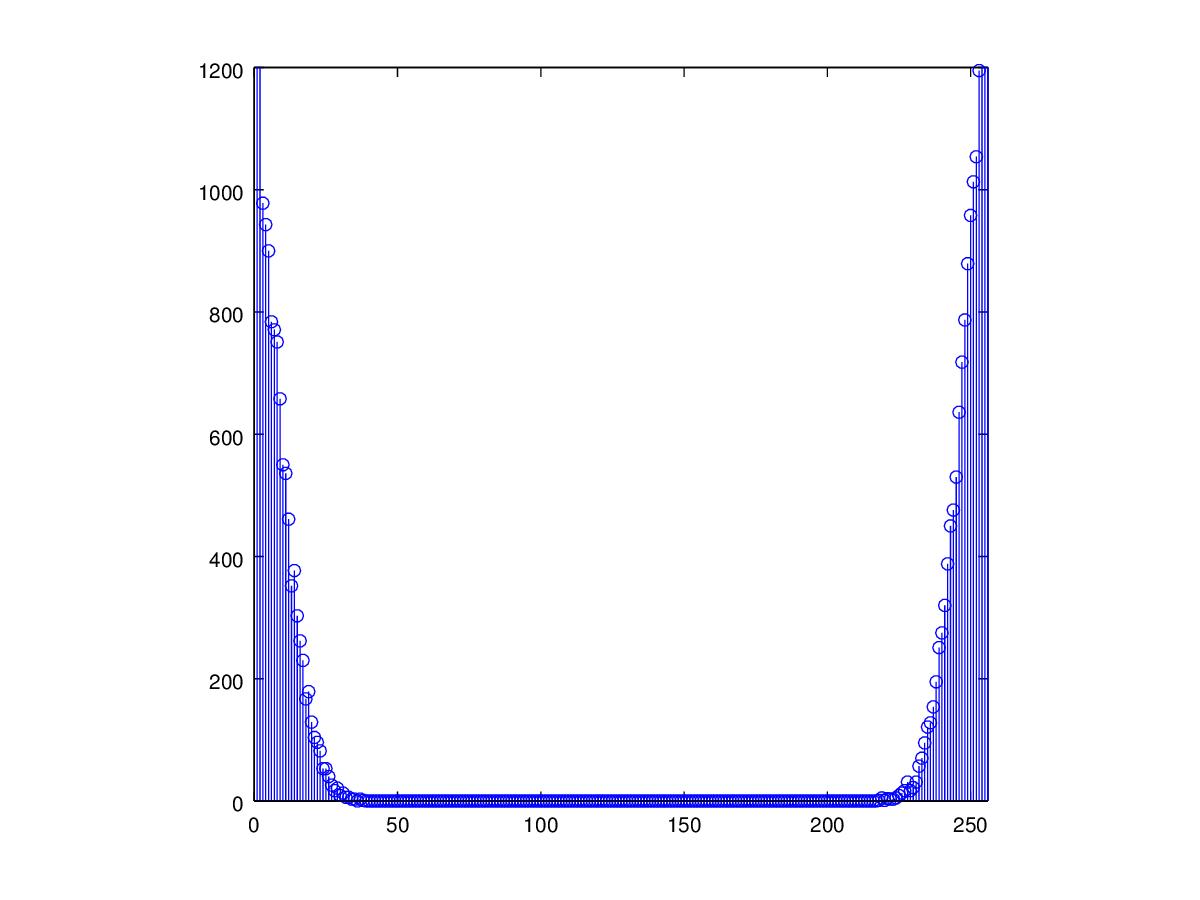}
 \includegraphics[height=2.975cm]{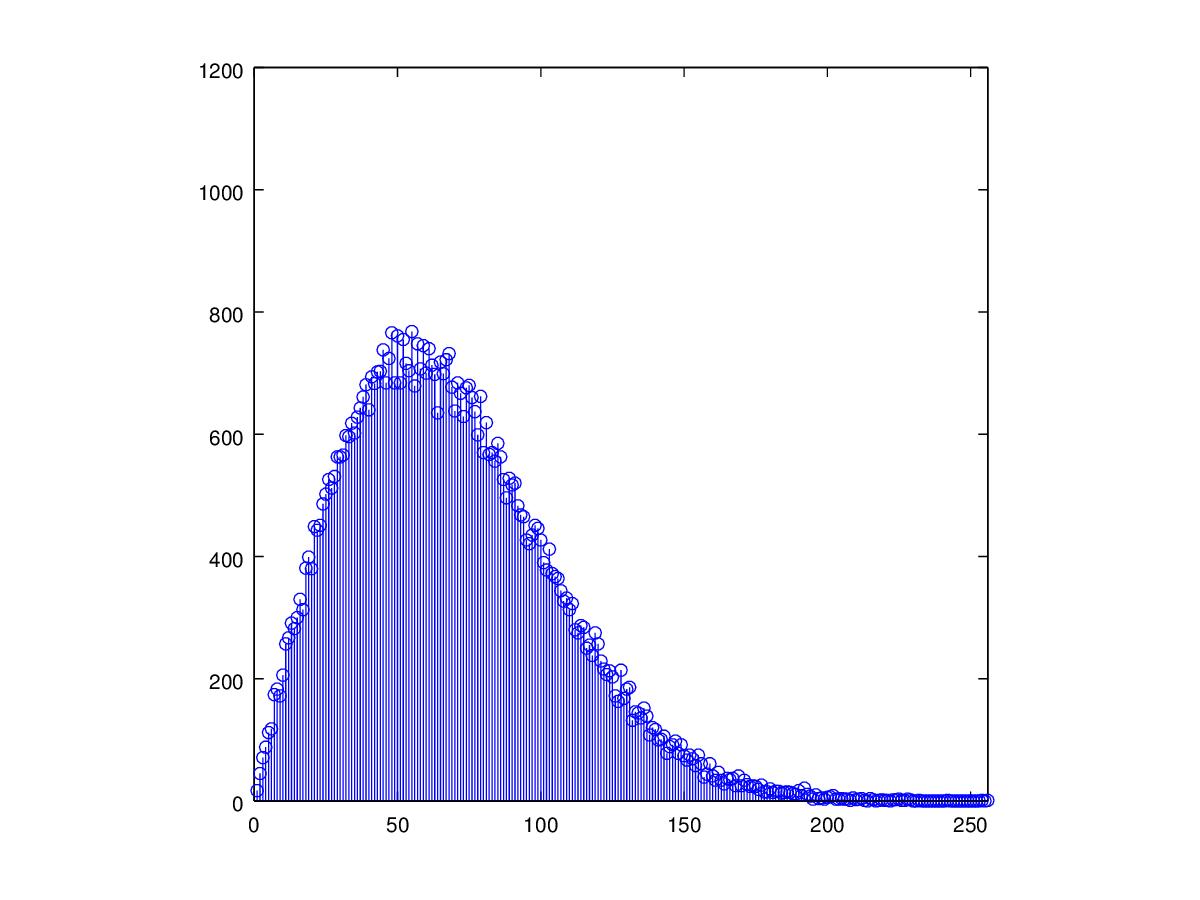} \\[5pt]
 \hglue -0.4in (e) \hskip 1.4in (f) \hskip 1.4in (g) \hskip 1.3in (h)\\[10pt]
 \includegraphics[height=2.975cm]{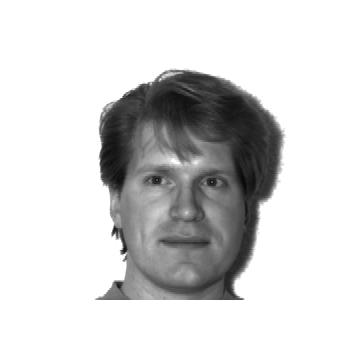}
 \includegraphics[height=2.975cm]{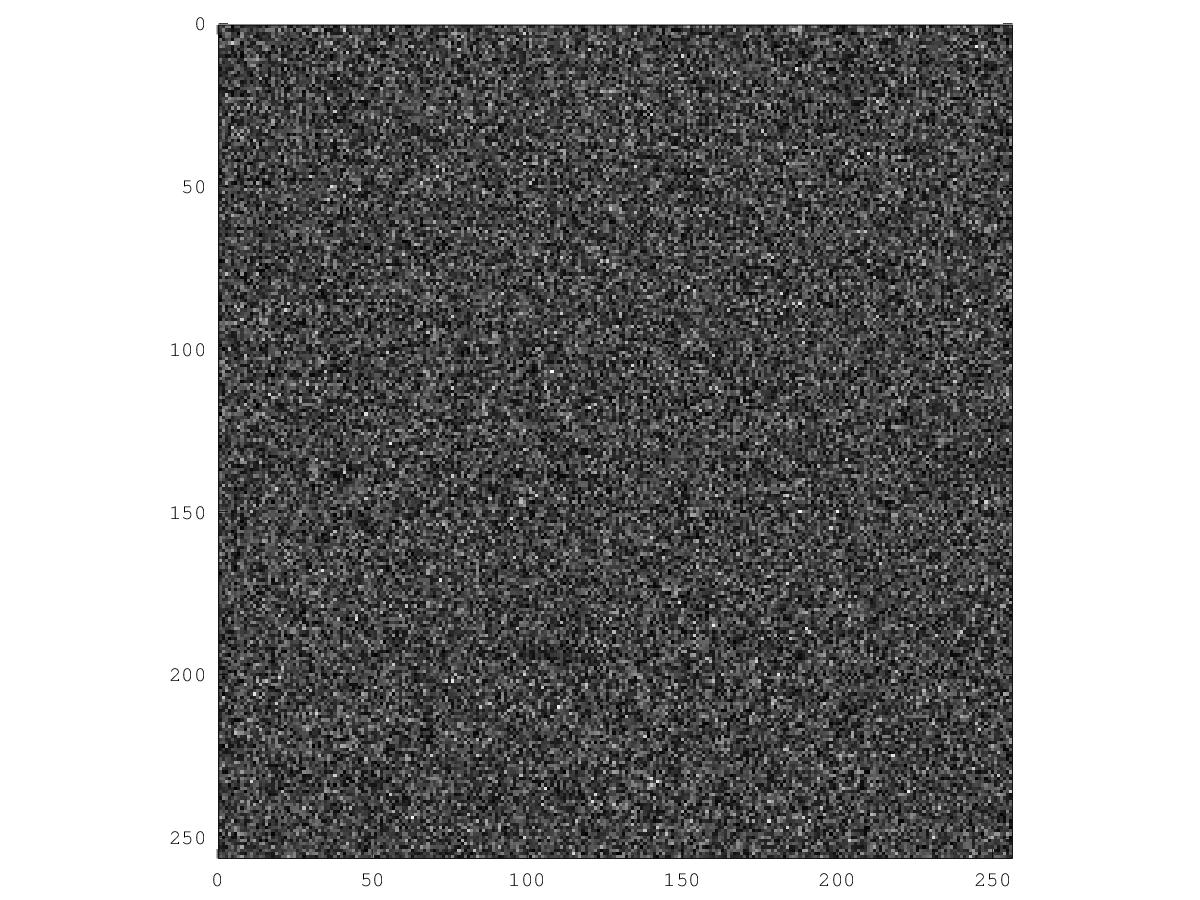}
 \includegraphics[height=2.975cm]{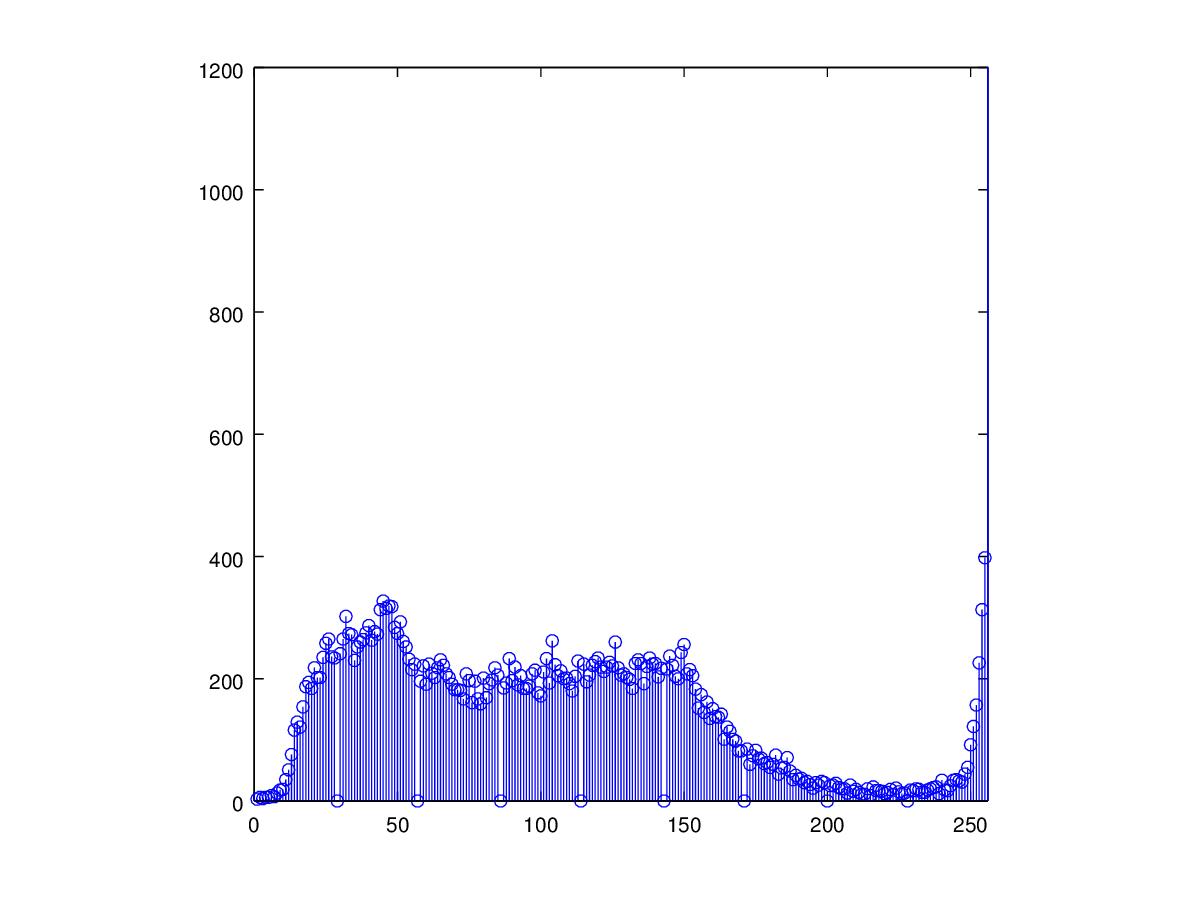}
 \includegraphics[height=2.975cm]{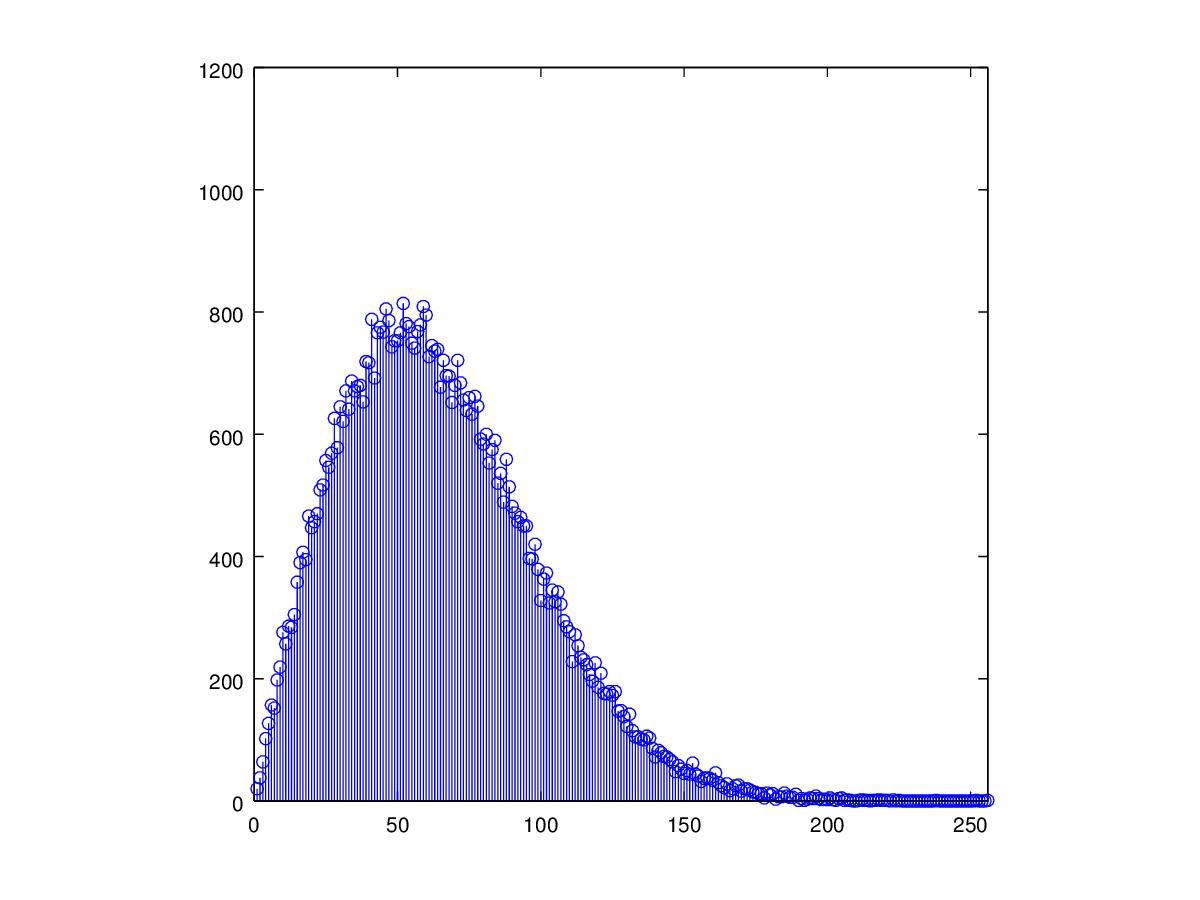} \\[5pt]
 \hglue -0.4in (i) \hskip 1.4in (j) \hskip 1.4in (k) \hskip 1.3in (l)\\[10pt]
 \caption{Results: (a)~fingerprint of right thumb, (e)~fingerprint of 
left thumb, (i)~face of the human subject; (b), (f) cipher images of the 
fingerprints by DRPE, (j) final privacy protected cipher image of the 
human face; (c), (g), (k) histograms of the plain images; (d), (h), (l) 
histograms of the cipher images}
  \label{results}
 \end{figure}

The feasibility of the proposed cryptosystem is tested by implementing 
the algorithm on image data of human face and fingerprints. To check our 
algorithm, we have used a few face images from the Yale Face Database B 
and have associated a few fingerprint images from the Bologna 
fingerprint data base. The face image is used as a plain text image, and 
the two fingerprint images are encrypted by DRPE method and the 
encrypted versions of the fingerprint images are utilized as encryption 
keys to encrypt the face image. The fingerprint images of a human 
subject's right and left thumbs, and a human subject's face are 
respectively shown in Fig.\ref{results}(a), \ref{results}(e) and 
\ref{results}(i) respectively, which are the plain images in our 
simulation experiment.
 The corresponding cipher images of the fingerprint images by DRPE are 
shown in Fig.\ref{results}(b) and (f). These encrypted images of 
Fig.\ref{results}(b) and (f) are utilized to produce the final privacy 
protected biometric signature PPBS by our proposed encryption algorithm 
and is shown in Fig.\ref{results}(j). The histograms of the plain images 
are shown in Fig.\ref{results}(c), (g) and (k) respectively. The 
histograms of the cipher images are shown in Fig.\ref{results}(d), (h) 
and (l) respectively. It is noticed from the results that there remains 
significant differences between the histograms of the plain images and 
that of the cipher images. According to the theory of classical 
cryptology it is good because this can be considered as diffusion, which 
diffuses the statistical characteristics of original image to the whole 
space. In addition, we also find that the cipher image of different 
original images have similar histograms after encrypted with independent 
encryption keys. This can be considered as confusion in the classical 
cryptology. It is difficult for attackers to obtain useful information 
from the statistical characteristics, thus it can provide considerable 
capacity of resisting statistical cryptanalysis.

The final privacy protected biometric signature can be decrypted by 
phase truncated Fourier transform algorithm of Fig.\ref{our_algo} using 
appropriate private phase keys PK$_1$ and PK$_2$. This result is shown in 
 \begin{figure}[htb]
 \centering
 \includegraphics[height=2.975cm]{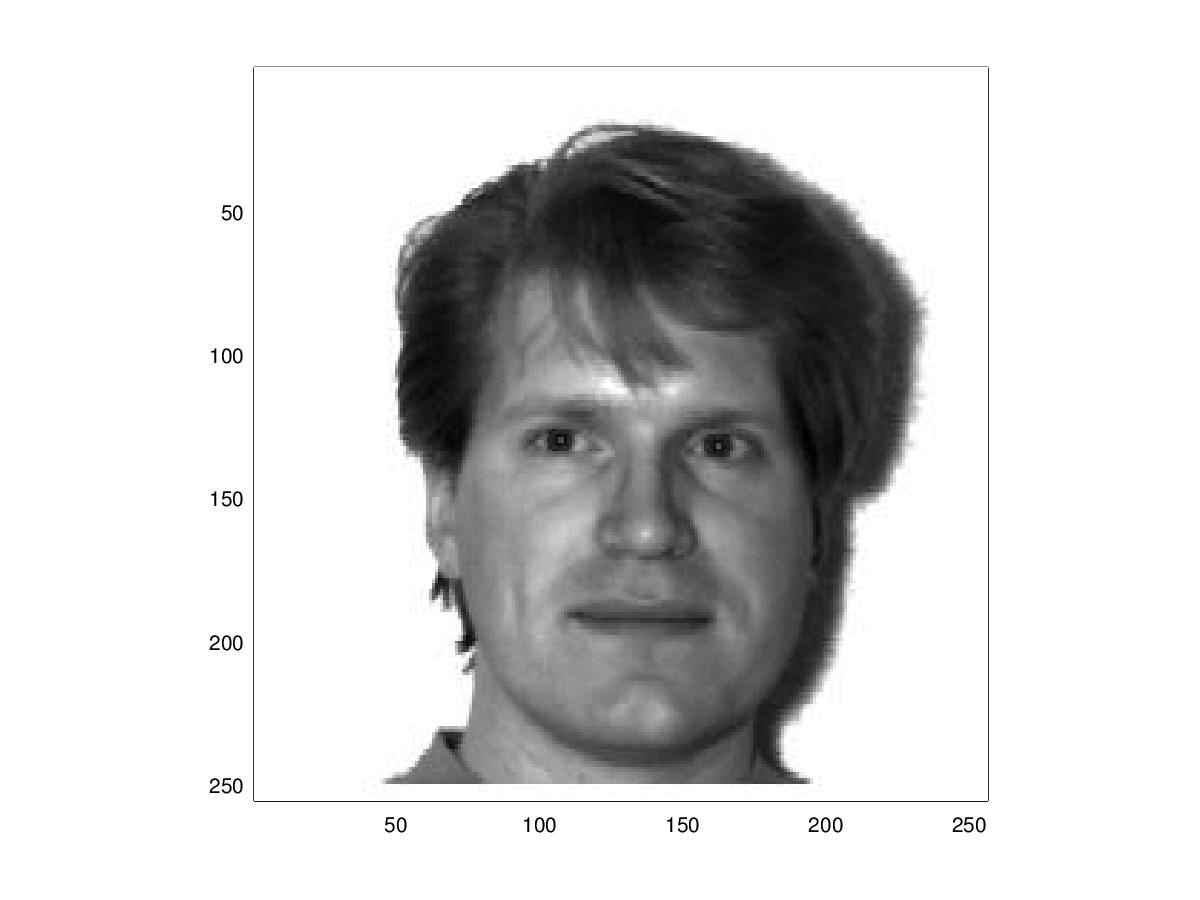}
 \includegraphics[height=2.975cm]{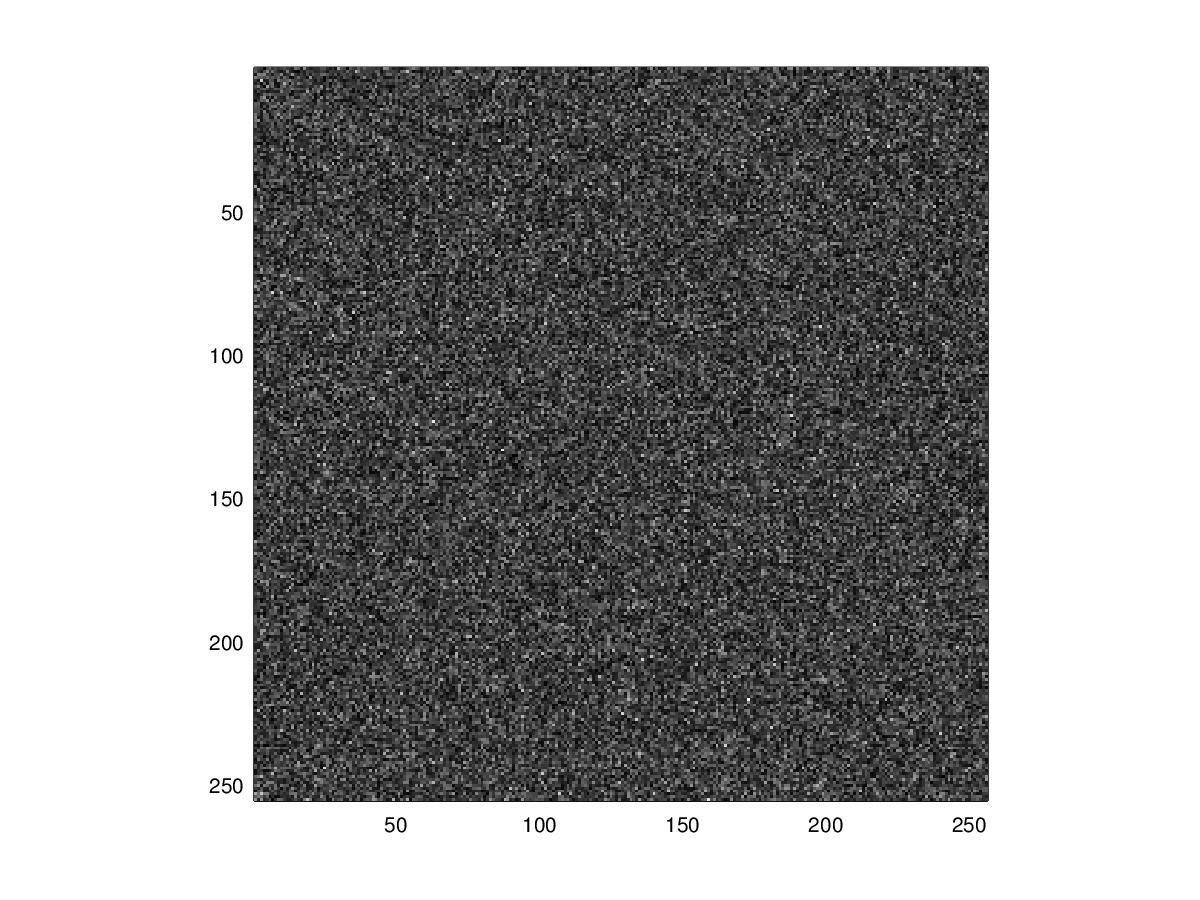} \\[5pt]
 (a) \hskip 1.4in (b)\\[10pt]
 \caption{Decryption result: (a)~decrypted image with correct phase 
keys, (b)~no decryption with a false key.}
  \label{result_check}
 \end{figure}
 Fig.\ref{result_check}. With correct phase keys, the original plain 
image of the human subject is extracted as shown in 
Fig.\ref{result_check}. When one of the phase keys, here the second 
phase key PK$_2$, is put incorrect, the algorithm produces an incorrect 
extraction which is a random image instead of the original face image as 
shown in Fig.\ref{result_check}(b).

\begin{figure}[htb]
 \centering
 \includegraphics[width=5.5cm]{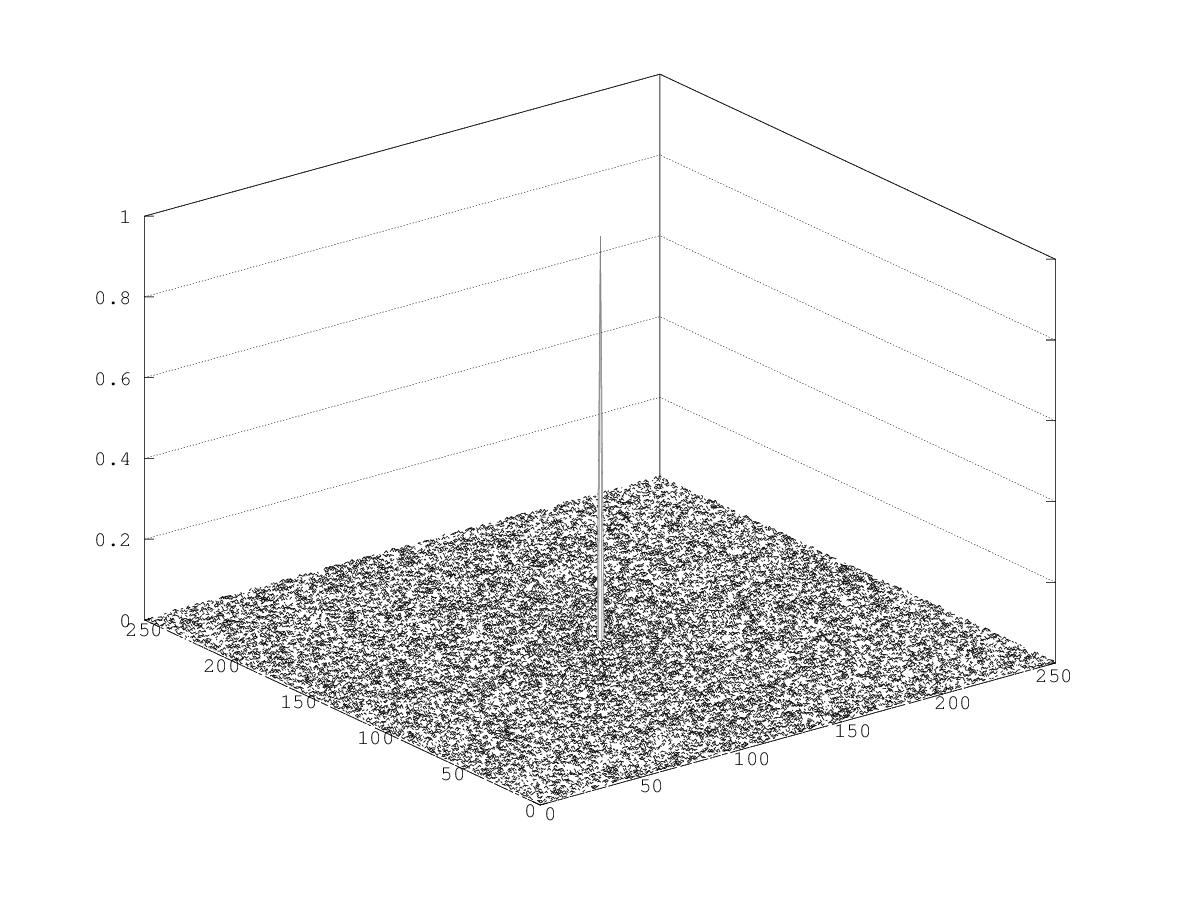}  \hskip 1cm
 \includegraphics[width=5.5cm]{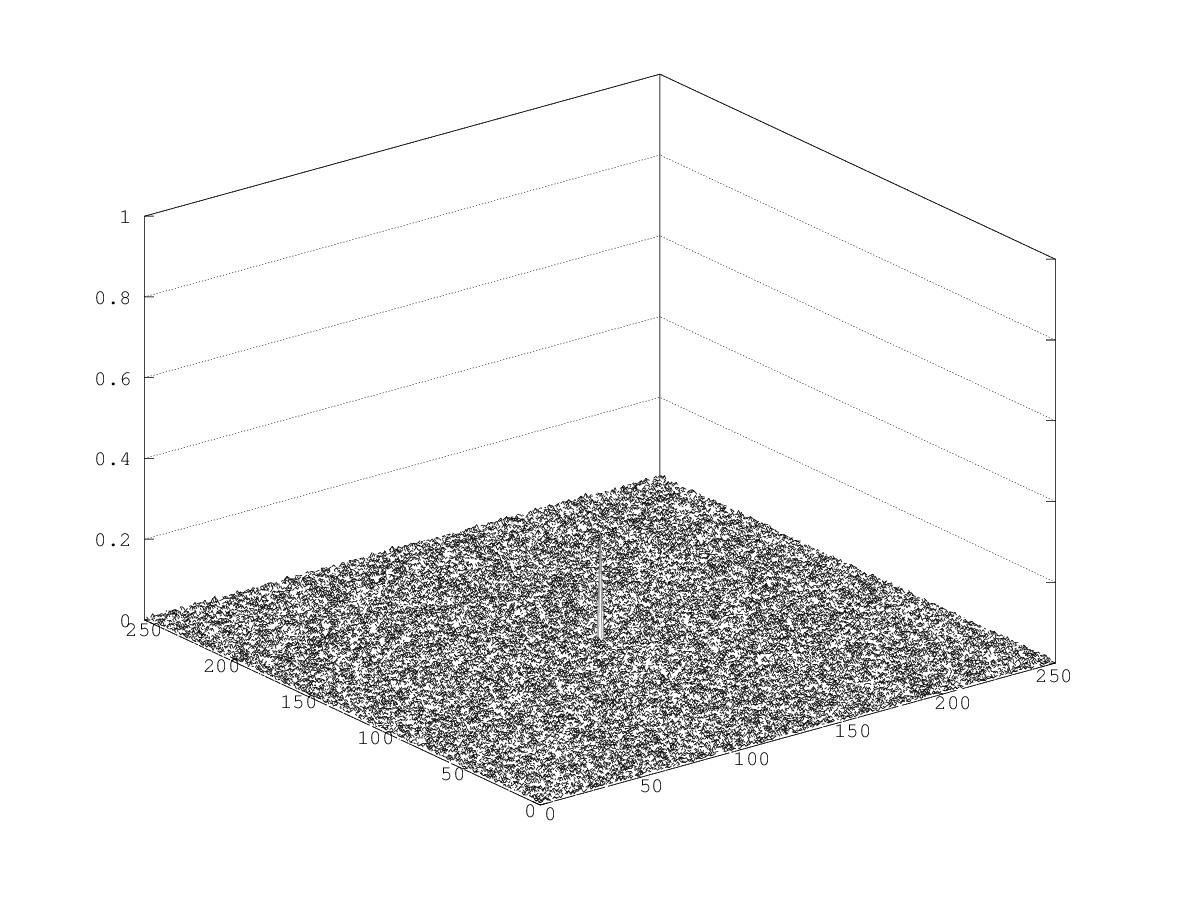} \\[5pt]
 (a) \hspace{5.8cm} (b) \\[10pt]
 \caption{Plots of cross-correlation: (a)~authentication of correct 
cipher biometric signature (correlation peak height=0.99), (b)~biometric 
rejection, no authentication (correlation peak height=0.2).  
\label{corr} }
 \end{figure}

The final encrypted PPBS can be used for authentication using 
cross-correlation~\cite{vanderlugt}. A high correlation peak 
authenticates the untampered cipher image and confirms no 
man-in-the-middle attacks. The height of the correlation peak determines 
whether it is a correct recognition or otherwise. Figure~\ref{corr} 
 shows the three-dimensional plots of the cross-correlation functions. A 
successful biometric recognition is confirmed by the high correlation 
peak of Fig.\ref{corr}(a) which is 0.99. When the PPBS of 
Fig.\ref{results}(j) is cross-correlated with a false signature, such as 
that of Fig.\ref{result_check}(b), a very low correlation peak is 
produced as shown in Fig.\ref{corr}(b) (which is equal to 0.2), and it 
signifies biometric rejection. All the implementation of the algorithms 
have been carried out using GNU Octave~\cite{gnu_octave}.

\section{Conclusion}

An asymmetric cryptosystem is proposed for privacy protection in 
biometric pattern recognition. The face biometrics is encrypted by 
fingerprint biometrics and double random phase encoding. Two private 
phase keys are generated by phase truncated Fourier transform. The 
encrypted final biometric image is privacy protected. The encrypted 
biometric signature is decrypted by using the synthesized two private 
phase keys. The encryption keys and the decryption keys are different. 
Authentication of the encrypted biometric signature is also carried out 
by cross-correlation operation. Simulation results verify the 
feasibility of the proposed cryptosystem.

\end{document}